\documentclass[prl,twocolumn,superscriptaddress,longbibliography]{revtex4-2}
\usepackage{amsmath,mathtools}
\usepackage[table]{xcolor}
\usepackage{subfigure}
\usepackage{ulem}
\usepackage{xcolor}
\usepackage{tabularray}
\usepackage{verbatim}
\usepackage{hyperref}
\usepackage{todonotes}
\usepackage{physics}

\newcommand{\mi}{\mathrm{i}}
\newcommand{\fleq}[1]{Eq.~\eqref{#1}}
\begin{document}

\title{ Cavity-QED tools for MBQC with optical binomial-codes}

\author{G.~P.~Teja}
\email[]{phaniteja.godavarthi@upol.cz}
\affiliation{Department of Optics, Faculty of Science, Palacký University, 771 46 Olomouc, Czech Republic}
\author{Radim Filip}
\email[]{filip@optics.upol.cz}
\affiliation{Department of Optics, Faculty of Science, Palacký University, 771 46 Olomouc, Czech Republic}

\begin{abstract}
Measurement-based quantum computation (MBQC) offers a promising paradigm for photonic quantum computing, but its implementation requires the generation of specific non-Gaussian resource states. 
While continuous-variable encodings such as the highly complex (GKP) states have been widely studied, the much simpler binomial codes offer an experimentally accessible alternative, though they demand a distinct set of operational tools. 
Here, we present a toolkit for MBQC using optical binomial codes, detailing a cavity-QED protocol for conditional generation of cluster states and the implementation of Pauli measurements. Our work proposes the first steps for existing optical atom-cavity architectures to lay the groundwork for their use in quantum computation.
\end{abstract}
\maketitle

Quantum computation possesses the remarkable potential to surpass classical computers in specific computational tasks. In recent years, the field has matured significantly, with prototype processors with 100 superconducting (SC) qubits \cite{2025_Jiang} while atomic qubit arrays have reached sizes of over 6000 qubits \cite{2025_Manetsch}.
Most quantum computers employ the circuit model. In this model, a register of qubits are initialized in a specific state, then a sequence of gates are applied to implement a desired unitary operation, and a final measurement yields the computational result.

An alternative to the circuit model is measurement-based quantum computation (MBQC), pioneered by the concept of the one-way quantum computer \cite{2001_rauss}. In MBQC, a specific entangled resource state, called a cluster state, is prepared initially. Computation then proceeds via adaptive local Pauli measurements, which expends entanglement to propagate information. These properties make photons natural candidates for MBQC. 
The universality of quantum computation that any unitary operation can be decomposed into a sequence of one and two-qubit gates \cite{1995_Barenco} holds equally in the measurement-based model. In fact, any unitary can be mapped to measurement patterns of CNOT and single-qubit gates on a 2D cluster state  \cite{2003_rauss}. Although 2D cluster states are sufficient for universal computation \cite{Larsen_2021}, 3D cluster states offer practical advantages, with higher error correction thresholds for fault-tolerant quantum computation (FTQC) \cite{2007_rauss,2010_Barrett}.

The principles of MBQC have been demonstrated across various platforms, including superconducting qubits \cite{Ferreira2024,2025_Sullivan}, atomic systems \cite{2022_Bluvstein}, and ion traps \cite{2013_Lanyon,Matsos2025}. Unlike these static qubit architectures, photonic qubits are inherently dynamic, and their detection inherently absorbs them. Photonic qubits can be encoded in various degrees of freedom, such as polarization, path encoding, or in continuous-variable states like GKP qubits \cite{2024_Shunya,Larsen2025,Eickbusch_2022,Matsos2025}. Among these, complex GKP states have emerged as particularly promising for MBQC, as cluster states can be generated using passive linear-optical elements and Pauli measurements are natively implemented via homodyne detection \cite{Larsen_2021}.

In this work, we investigate much simpler binomial codes as an alternative encoding for photonic measurement-based quantum computation (MBQC). These codes, being finite superpositions of low photon number states, offer a more experimentally accessible platform at optical frequencies compared to intricate GKP codes that require large photon number superpositions. We present the essential tools for this approach, comprising the conditional generation of optical binomial code states, deterministic implementation of a CZ-gate, deterministic construction of cluster states, and ancilla-based Pauli measurements. These four validation steps are key and closest milestones for upcoming experimental tests.
Our effort is stimulated by progress in superconducting platforms, where binomial codes have already been successfully demonstrated \cite{2019_Hu}. In superconducting systems, the availability of strong dispersive interactions \cite{Eickbusch_2022, 2024_Landgraf} and nonlinearities \cite{Eriksson_2024,laha2025} enables complete control over bosonic modes. Furthermore, the existence of high-fidelity CNOT gates \cite{Chou_2018, Rosenblum_2018} makes the circuit model of computation and error correction straightforward to implement \cite{Ma_2020}.

However, implementing binomial codes in optical pulses remains an open problem, although they are low-dimensional Fock superpositions, as they lack Hamiltonian flexibility \cite{Eickbusch_2022, 2024_Landgraf, Matsos2025} and are limited primarily to Jaynes-Cummings (JC) type interactions. Therefore, proposing binomial codes and developing tools for MBQC is essential for experimental work in this unexplored direction of quantum optical computing. We note that further steps in error detection and correction, while critical for scalable quantum computing, are beyond the scope of this initial work.

The Knill-Laflamme condition, that guarantees correctable protection against set of errors
$
\mathcal{E} \equiv \qty{I,~\hat{a}\dots\hat{a}^L, ~\hat{a}^\dagger\dots\hat{a}^{\dagger G},~\hat{n}\dots \hat{n}^{D}},
$
can be stated as:
\begin{align}\label{eq:kl}
\mel{W_0}{\hat{E}_k^\dagger \hat{E}_l}{W_1} = \alpha_{kl} \delta_{01},
\end{align}
where $\alpha_{kl}$ is a hermitian matrix.
$\hat{E}_k \in \mathcal{E}$ and $W_{0/1}$ are code states that can be noise corrected against noises in $\mathcal{E}$. 
It has been shown that a large class of states known as binomial code state satisfy the condition \fleq{eq:kl} for arbitrary polynomials of creation/annihilation operators \cite{Michael_2016}. Binomial code states are written as
\begin{align}\label{eq:cod}
\ket{\tilde{0}/\tilde{1}} = \dfrac{1}{\sqrt{2^N}} \sum_{p~\text{even/odd}}^{[0,N+1]}
\sqrt{N+1 \choose p} \ket{p(s+1)}.
\end{align}
The binomial codes for photon-loss errors \cite{Michael_2016} i.e. $\mathcal{E}_1 = \qty{I,\hat{a}}$, are written as \fleq{eq:cod}:
\begin{align}\label{eq:c4}
\ket{\tilde{0}} = \dfrac{ \ket{0}+\ket{4} }{\sqrt{2}}, \quad \ket{\tilde{1}} = \ket{2} 
\end{align}
and similarly the dephasing loss $  \mathcal{E}_2 = \qty{I,\hat{a}, \hat{n}} $ can be corrected using higher Fock states in \fleq{eq:cod} or by entangling the code states of $\mathcal{E}_1$ in two modes \fleq{eq:c4}
\begin{subequations} \label{eq:cd2}
\begin{align}
\ket{\tilde{0}} &= \dfrac{ \ket{0}+\sqrt{3}\ket{4} }{2}, \quad \ket{\tilde{1}} = \dfrac{ \sqrt{3} \ket{2}+\ket{6} }{2},  \\
\ket{\tilde{0}} &= \dfrac{ \ket{04}+\ket{40} }{\sqrt{2}}, ~\quad   \ket{\tilde{1}} = \ket{22}, 
\end{align}
\end{subequations}

Our protocol enables the preparation of states in \fleq{eq:c4} and can be extended to generate states in \fleq{eq:cd2}. 
Since photon loss remains the dominant error channel for optical codes, we focus on the lowest-order binomial codes given in Eq.~\eqref{eq:c4} and use the fidelity
$
\mathcal{F}(\rho_1,\rho_2) = \qty( \text{tr} \sqrt{ \sqrt{\rho_1} \rho_2 \sqrt{\rho_1} } ) ^2
$
to quantify state quality.
Unlike GKP codes, which exhibit translation symmetry, binomial codes possess rotation symmetry: the operator $e^{i \frac{\pi}{2} \hat{n}}$ acts as a logical $\sigma_z$ in the subspace ${\ket{\tilde{0}}, \ket{\tilde{1}}}$.

\begin{figure}[!tbh] \label{fg:pst}
	\subfigure[Binomial-code generation \label{cz0}] {\raisebox{0.3cm}{\includegraphics[scale=0.62]{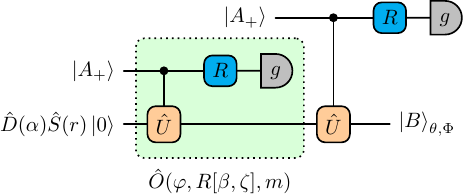}}} \hfill
	\subfigure[Density matrix \label{den}] {\includegraphics[scale=0.32]{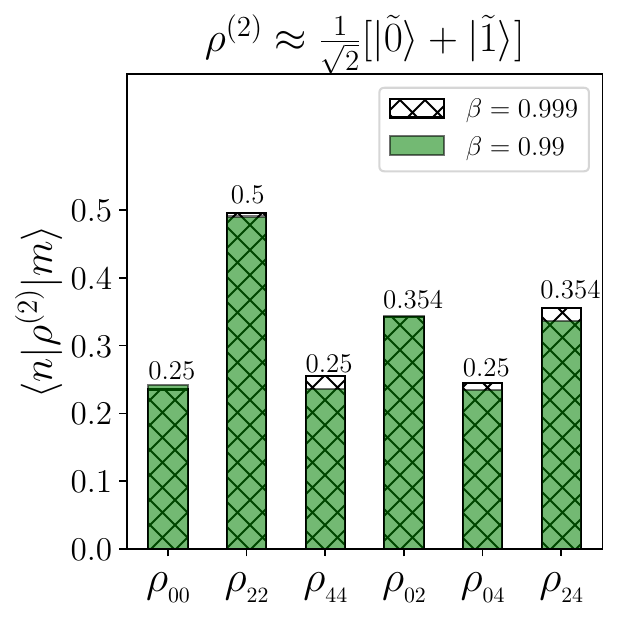}}
\caption{\subref{cz0} Preparing superposition states of binomial codes using Gaussian inputs, $\hat{O}(\varphi,R,m)$ denotes a single atom-cavity iteration \fleq{eq:ite}. 
\subref{den} $\ket{+}$ state is generated using the circuit in \subref{cz0}, other density matrix elements are negligible. $\beta$ is the cavity efficiency. A single iteration $\hat{O}(\pi, H, m)$ using cavity QED is demonstrated in \cite{Hacker_2019} to generate cat states.}
\end{figure}

\paragraph{Cavity QED interaction:}
We follow the experimental protocol for optical cat-state generation using cavity QED \cite{Hacker_2019} to conditionally filter binomial code superpositions from Gaussian inputs. This is achieved using a controlled-phase-flip (CPF) operation $U(\varphi)= e^{i \varphi \hat{n}} \otimes \dyad{g} + I \otimes \dyad{s}$, followed by an atomic rotation $R$ and an atomic measurement.
The crucial component of the toolkit is the CPF realized via  atom-cavity reflections and we account for the scattering losses using Kraus operator framework~\cite{2022_hastrup}, avoiding  virtual cavity methods \cite{2019_klaus,2020_kii,teja2023} which are challenging for multiple reflections.
The density matrix evolves as $\mathcal{E}_\varphi (\rho) \to \sum_{j} K_j (\varphi) \rho K_j^\dagger (\varphi)$, where $K_j (\varphi)$ adds noise to the $U(\varphi)$ operation. This noise depends on the cooperativity $C=g^2/(\kappa\gamma)$ and cavity efficiency $\beta=\kappa_c/\kappa$ ($\kappa \equiv \kappa_c + \kappa_l$), where $g$, $\gamma$, $\kappa_c$, and $\kappa_l$ denote the atom-cavity coupling, atomic decay, cavity emission, and loss rates, respectively.
Numerical simulations are performed using QuTiP \cite{qutip5}. The atom-light state after the three operations transforms as:
\begin{subequations}
\begin{align}
\hat{O}(\varphi,R,m)  &\equiv \dyad{m} \otimes R[\alpha,\beta,\zeta] \otimes \mathcal{E}_\varphi (\rho)  \label{eq:ite}\\
\rho^{(n+1)} &\equiv \hat{O} \,\mathcal{E}_\varphi (\rho^{(n)})\, \hat{O}^\dagger/\text{Tr}[\hat{O}\,\mathcal{E}_\varphi (\rho^{(n)})\,\hat{O}^\dagger],
\end{align}
\end{subequations}
here, $\ket{m} \in \{\ket{g}, \ket{s}\}$, $R$ is unitary matrix \fleq{app:unit} and $\rho^{(n)}$ is the state of light mode after $n$ atom cavity iterations.
The CPF between an atomic qubit and an optical mode is realized via standard JC-interactions \cite{Tiecke_2014,Hacker_2016,Staunstrup2024}. 
We start by conditionally preparing code and magic states from displaced squeezed vacuum states. The input state and the superposition of binomial codes are defined  as:
\begin{align}
D(\alpha)S(r)\ket{0} = &\sum_{n=0}^{5} c_n \ket{n} + \mathcal{O}^{(6)},\\
\ket{\mathrm{B}}_{(\theta,\Phi)} = & \cos{\theta} \ket{\tilde{0}} + e^{\mi\Phi} \sin{\theta} \ket{\tilde{1}}, \label{eq:bst}
\end{align}
note that a photon number dependent phase $\Phi$ cannot be achieved by optimizing $\qty{\alpha, r}$ and is typically achieved in SC platforms using SNAP-gates \cite{2024_Landgraf}. Such SNAP gates are not yet available for optical light. 
To generate general superposition states, we first set $\Phi=0$ and optimize $\alpha$ and $r$ to approximate binomial superpositions with arbitrary $\theta$, satisfying:
\begin{itemize}
	\item Negligible high Fock-state contributions, $\sum_{n=0}^5 \abs{c_n}^2 \approx 1$.
	\item Matching the target amplitude ratios $c_{\tilde{1}}/c_{\tilde{0}} = \tan \theta$  i.e. $c_2/c_{\{4,0\}} = \sqrt{2}\tan\theta$ and $c_4/c_0 = 1$.
\end{itemize}
Figs.~\ref{p4}-\ref{p33} reveals only $\theta = \pi/3.3$ is directly attainable. However, starting from $\ket{\mathrm{B}}_{\pi/3.3,0}$, a second atom-cavity iteration with $\hat{O}(\pi/2, R[\beta,\zeta], g)$ produces general binomial superposition states where $\theta, \Phi$ are controlled by atomic rotation (SM):
\begin{align} \label{eq:arn}
	\frac{-\cos\zeta+ e^{\mi \beta}\sin\zeta}{\cos\zeta + e^{\mi \beta}\sin\zeta} \tan\left(\frac{\pi}{3.3}\right) = \tan\theta~e^{i\Phi},
\end{align}
\fleq{eq:arn} admits numerical solutions for diverse target states (Fig.\ref{fg:ap1}). For instance, $\qty{\zeta,\beta} = \qty{-1.714,0}$ yields $\ket{\mathrm{B}}_{\pi/4,0}$, producing the $\ket{+}$ state. Further optimization of the second rotation enables generation of $T$-type and $H$-type magic states, which are essential resources for non-Clifford gate teleportation \cite{Bravyi_2005}. The preparation of general superposition states is summarized in Fig.~\ref{cz0}.

Atomic measurements are profitable for conditionally generating high-fidelity target states. Tracing out the atomic system after first and second reflections yields the target state $\ket{+}$ with fidelities (0.50, 0.25), while projective measurements achieve higher fidelities (0.97, 0.98). Further, the second measurement is also indispensable, although the state after first reflection  $(\ket{\mathrm{B}}_{\pi/3.3,0})$ approaches the target state $\ket{+}$, feed forward Gaussian operations cannot improve its fidelity, see SM Fig.~\ref{fg:feed}.

\begin{table}[!tbh]
\renewcommand{\arraystretch}{1.2} 
\setlength{\tabcolsep}{6pt}
\begin{tabular}{w{l}{0.8cm}|l|>{\columncolor{gray!20}}r|c|>{\columncolor{gray!20}}c|c|>{\columncolor{gray!20}}c}
 Target &	\multicolumn{1}{c}{$\mathcal{F}^{(1)}_{0.99}$} & $\mathcal{F}^{(2)}_{0.99}$ & \multicolumn{1}{c} {$\mathcal{F}^{(1)}_{0.999}$} & $\mathcal{F}^{(2)}_{0.999}$ & $\mathcal{M}$& $\mathcal{F}_T$\\
\colrule
$\ket{+}$ & 0.950 &  0.964 & 0.968 &  0.985 & 0.23 & - \\ 
$\ket{T_1}$ & 0.815 & 0.954 & 0.830 & 0.985 & 0.23 & 0.910\\
$\ket{T_2}$ & 0.663 & 0.942 & 0.676 & 0.983 & 0.19& 0.910 \\ 
$\ket{H}$ & 0.702 & 0.947 & 0.716 & 0.983 & 0.18 & 0.927\\
$\ket{\mathcal{A}}_{{\pi}/{3}}$  & 0.746 & 0.953 & 0.760 & 0.988 & 0.24&-
\end{tabular}
\caption{\label{tab:mes}%
$\mathcal{F}^{(n)}_\beta$ denotes the fidelity of states ($\ket{+},\ket{T_1}, \ket{T_2}, \ket{H} ,\ket{\mathcal{A}}_{{\pi}/{3}}$) after the $n^\text{th}$ atom-cavity iteration with scattering losses $\beta$. $\mathcal{M}$ represents the atomic measurement success rate, which remains similar for both $\beta$ values $\qty{0.99,0.999}$. $\mathcal{F}_T$ denotes the threshold fidelities of magic states required for fault-tolerant computation \cite{Bravyi_2005}. For definitions of Magic states ($\ket{T_1}, \ket{T_2}, \ket{H}$) and optimized values of atomic rotation see SM (Table.~\ref{tab:sol}).  $\ket{\mathcal{A}}_{t}$ is the ancilla state required for POVM \fleq{eq:pvm}.}
\end{table}

As discussed earlier, the generation of binomial superposition states (\fleq{eq:bst}) follows the circuit in Fig.~\ref{cz0}, beginning with a displaced squeezed state $\ket{1.4,0.25}$ in the optical mode. First atom-cavity iteration $\hat{O}(\pi/2,H,g)$ remains identical for all target states, while the atomic rotation in the second iteration is optimized to produce the states in Table.~\ref{tab:mes}.
The density matrix elements of the resulting $\ket{+}$-state (Fig.~\ref{den}) show well-preserved coherences $\rho_{mn} = \mel{m}{\rho}{n}$, consistent with high state fidelity. The protocol's success probability is numerically obtained to vary between 0.18–0.23 for two atomic measurements and remains similar across different scattering losses. Notably, our protocol generates $T$-type and $H$-type magic states with fidelities exceeding 0.98, surpassing the 0.927 threshold required for universal quantum computation \cite{Bravyi_2005}, establishing a viable path towards the first experimental milestone in quantum computation with optical binomial codes.

\begin{figure}[!tbh]\label{fg:czgat}
	\subfigure[Deterministic CZ-gate \label{cz}] {\raisebox{0.5cm}{\includegraphics[scale=0.66]{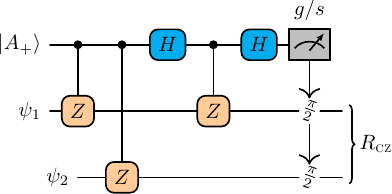}}}	\hfill
	\subfigure[$ \Delta R_\text{CZ} $  ($\beta = 0.999$) \label{cz1}] {{\includegraphics[scale=0.32]{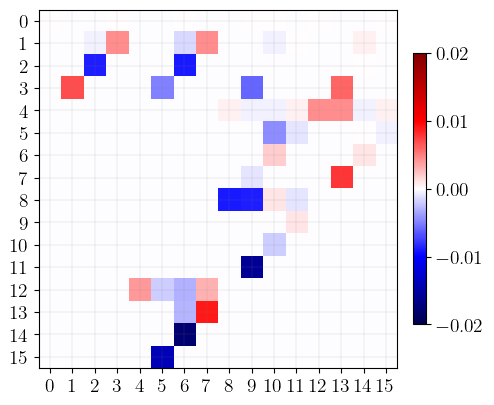}}}
	\caption{CZ-gate between two optical modes encoded in binomial code states. 
		\subref{cz} Implementation of the CZ-gate between optical modes; the feed-forward phase shift $e^{i\frac{\pi}{2}\hat{n}}$ is applied only for $\ket{s}$ measurement outcomes. Here, the atom-photon CZ-gate is implemented as a single iteration $\hat{O}(\frac{\pi}{2}, I, I)$, without the atomic rotation and projective measurement.
		\subref{cz1} Difference between the process map and the ideal CZ-gate  ($\Delta R_\text{CZ}$). For cavity loss $\beta= \qty{0.999, 0.99}$, $\max \abs{\Delta R_\text{CZ}} = \qty{ 0.018, 0.099}$ is obtained using \fleq{eq:dcz}. }
\end{figure}

\paragraph{Deterministic CZ-gate:}
Once the superposition of binomial code states is generated, deterministically obtaining the entangling CZ-gate is conceptually straightforward within cavity QED setups \cite{Rosenblum_2018}. The CPF $U(\pi/2)$ operation in the code basis $\{\ket{\tilde{0}}, \ket{\tilde{1}}\}$ yields  $ Z \dyad{g} + I \dyad{s}$. Initializing the atom in $\ket{A_+} = (\ket{g} + \ket{s})/\sqrt{2}$ and applying the circuit in Fig.~\ref{cz} gives the output state prior to atomic measurement as
\begin{align} 
	\ket{\Psi} =  \qty[\small{\text{diag(1,1,1,-1)}} \ket{g} + \small{\text{diag(1,-1,-1,-1)}} \ket{s} ] \ket{\mathrm{B}} \ket{\mathrm{B}},
\end{align}
where $\mathrm{diag}(\cdot)$ denotes a diagonal matrix in the basis $\qty( \ket{\tilde{0}\tilde{0}}, \ket{\tilde{0}\tilde{1}}, \ket{\tilde{1}\tilde{0}}, \ket{\tilde{1}\tilde{1}})$ and $\ket{\mathrm{B}}$ is the binomial superposition state (\fleq{eq:bst}). The deterministic CZ-gate is implemented by conditioning on atomic measurement outcomes: $\ket{g}$ measurement directly yields $R_\text{CZ} = \dyad{\tilde{0}} I + \dyad{\tilde{1}} Z $, while a $\ket{s}$ outcome requires a feed-forward phase shift $e^{i \pi \hat{n}/2}$ on the light modes to deterministically realize the $R_\text{CZ}$ operation. 

We characterize quality of the CZ operation via  process tomography \fleq{eq:dcz}, 
$
\rho_{\text{out}} = R_\text{CZ} \rho_{\text{in}}
$
where  $\rho_\text{in}$ is a complete set of input states (\fleq{eq:bss}). For $\beta = \qty{0.999, 0.99}$, the process map $R_\text{CZ}$ (Fig.~\ref{cz1}) exhibits a maximum deviation of $\qty{0.018,0.099}$ from the ideal case, which translates to fidelities of $\qty{0.98,0.93}$  (see SM Fig.~\ref{cz_st} for density matrix comparisons).
The experimental realization of a deterministic CZ-gate mediated by two atom-cavity reflections between two cavities constitutes the second crucial milestone for advancing quantum computation with binomial codes.

\begin{figure*}[!tbh]
 \flushleft	
	\subfigure[5-star cluster generation\label{cl5}] {\includegraphics[scale=0.6]{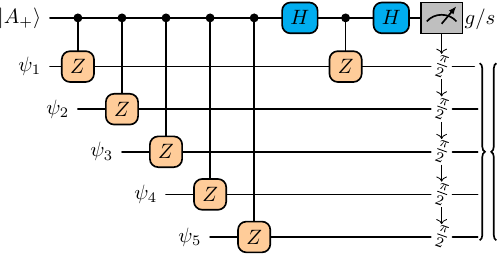}} 
	\subfigure[RHG lattice \label{cube}]{\raisebox{0.4cm}{\includegraphics[scale=0.55]{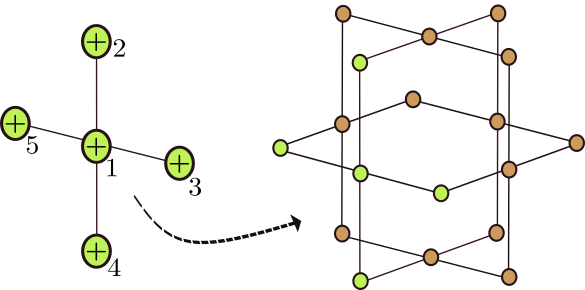}}} 
	\subfigure[Mean stabilizers \label{stb}] {\includegraphics[scale=0.29]{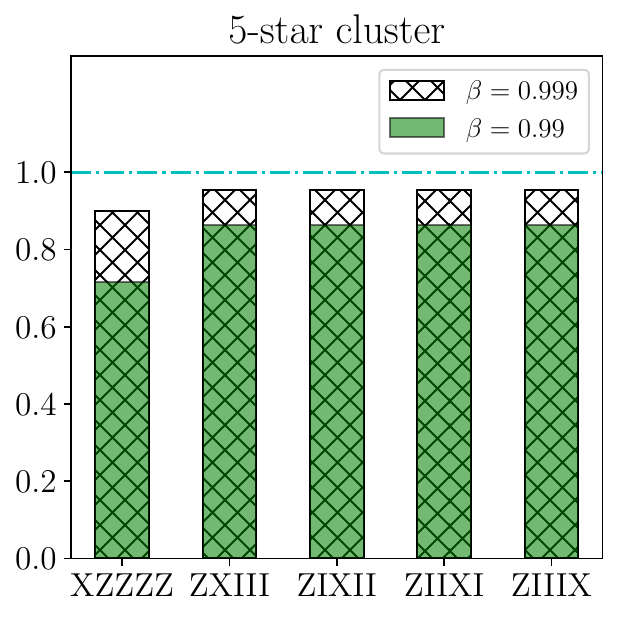}} \hfill
	\subfigure[POVM \label{mc}] {\includegraphics[scale=0.27]{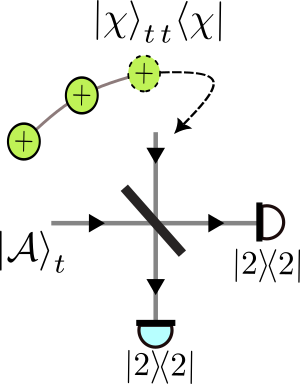}}
\subfigure[Fidelities \label{tab:sol1}]{\raisebox{1.3cm}{
	\renewcommand{\arraystretch}{1.5} 
	\setlength{\arraycolsep}{1.5pt} 
	\resizebox{0.15\textwidth}{!}{%
	\begin{tabular}{w{c}{1cm}|c|c|c}
					state &  $\mathcal{F}_{0.99}$  &  $\mathcal{F}_{0.999}$ & $\mathcal{P}$  \\
					\colrule \colrule
					3-chain	&   0.997 &  0.999 & 0.145  \\	
					5-star	&   0.997 &  0.999 & 0.145   \\
							
	\end{tabular}
	}}}
\caption{Deterministic generation of cluster states with binomial codes. 
\subref{cl5} 5-Star shaped cluster state generation, which can be foliated to construct a unit cell of the RHG lattice as shown in \subref{cube}. Vertices represent $\ket{+}$ states in Fig.~\ref{fg:pst} and solid lines denote CZ-gate in Fig.~\ref{fg:czgat}.
\subref{stb} Stabilizer measurements for the star-shaped cluster state generated using the circuit in \subref{cl5}.\subref{mc} Conditional projective measurement $\ket{\chi}_t$ in the XY-plane on a cluster graph using PNRD's and ancillary state. Table~\subref{tab:sol1} shows the fidelities of post-measurement states with respect to ideal projected states for both 3-chain \subref{cl5} and 5-star cluster states \subref{mc}. In both configurations, we take an ideal cluster state and project a qubit to $\ket{\chi}_{\pi/3}$, and similar results are observed varying the qubit location and the projection angle $t$.}
\end{figure*}

\paragraph{Binomial cluster states:}
Cluster states can be intuitively understood as graphs where vertices are initialized in the $\ket{+}$ state and edges are created by applying CZ-gates between selected pairs of light modes, as illustrated in Figs.~\ref{cl5} and~\ref{cube}. While a deterministic, high-fidelity CZ gate suffices to generate arbitrary cluster-state graphs, performing atomic measurements for each link in the  cluster state makes it resource-intensive.
We consider an optimized approach, shown in Fig.~\ref{cl5}, which requires only a single atomic measurement for a 5-star cluster state \cite{Ferreira2024}. Similar to $R_\text{CZ}$ map, a gate in Fig.~\ref{cl5} for cluster state generation is also deterministic, requiring only a single feed-forward phase shift only for $\ket{s}$ measurement outcomes. A star-shaped cluster state generated via this method can be foliated into a three-dimensional Raussendorf-Harrington-Goyal (RHG) lattice \cite{Raus_2007}, as shown in Fig.~\ref{cube}, which is a resource for fault-tolerant quantum computation (FTQC).
This path to FTQC is encouraged by recent demonstrations of 3D cluster states in the time-frequency modes of light \cite{Roh_2025} and the deterministic generation of 2D cluster states at optical \cite{Larsen_2021} and microwave frequencies \cite{2025_Sullivan}.
Another promising extension involves integrating propagating binomial codes into cavity-memory breeding architectures, where wave packets at different times are stored and then released for conditional detection, enabling the generation and entangling of non-classical states \cite{hanamura2025}.
Although cluster states are understood as entangled graphs, they are formally defined as the common eigenvector of stabilizer operators \cite{2001_rauss}:
\begin{align}
	S_i = X_i \prod_{j \in \textsc{n}_i} Z_j
	\quad \text{with} \quad 
	S_i \ket{\mathcal{C}} = \pm \ket{\mathcal{C}},
\end{align}
here, $X = \dyad{\tilde{0}}{\tilde{1}} + \dyad{\tilde{1}}{\tilde{0}}$ and $Z = \dyad{\tilde{0}} - \dyad{\tilde{1}}$ are Pauli operators in the binomial code basis, and $N_i$ denotes the set of neighboring vertices. The cluster state $\ket{\mathcal{C}}$ is the common eigenvector of these stabilizers.
. For instance, the star-shaped cluster state (Fig.~\ref{cl5}) is stabilized by the set $S = \qty{XZZZZ,~ZXZZZ, \dots, ZZZZX}$. To assess the quality of the generated cluster state, we evaluate the averages of these stabilizers rather than relying on fidelity measures \cite{Thomas_2024}. Figure~\ref{stb} shows the expectation values $\expval{S_i} {\mathcal{C}}$ for a cluster state produced by simulating the circuit in Fig.~\ref{cl5}, revealing that low cavity losses are crucial for maintaining the stabilizer values.

Furthermore, to evaluate the entanglement quality of cluster states \cite{2021_Qin}  we replace qubit 4 in Fig.~\ref{cl5} with the state $\ket{B}_{(\frac{\pi}{3},-\frac{\pi}{5})}$, then measure qubits 1-4 in the $\qty{XZZIX}$ basis, thereby teleporting the state to qubit 5. Numerical simulations give teleportation fidelities $\qty{0.98,0.96}$ for $\beta = \qty{0.999,0.99}$, which surpasses the threshold of 2/3 required for a reliable quantum channel \cite{2021_Qin}. Ultimately, scaling this architecture to generate 3D cluster states requires achieving the third experimental milestone of sequential interaction between multiple optical modes and an atom-cavity system.

In MBQC, computation proceeds through measurements in the XY-plane \cite{2001_rauss,mantri2017} i.e. in the eigen-basis of the operator, specifically $\cos t \hat{X} + \sin t \hat{Y}$:
$
\frac{1}{\sqrt{2}} \qty(\ket{\tilde{0}} \pm  e^{\mi t}  \ket{\tilde{1}}).
$
A conditional version of  measurements can be advantageously performed on binomial codes without requiring an inline atom during cavity-cavity interactions. We employ the scheme in Fig.~\ref{mc} with two photon-number-resolving detectors (PNRD's) \cite{2009_Garciana}, a beam-splitter (BS), and an ancillary state. The ancillary state is an optimized binomial superposition, which can be prepared using optimized atomic rotations as discussed above. The ancilla ($\ket{\mathcal{A}}_t$) and the corresponding POVM  $(\ket{\chi}_{t})$ for detecting two photons are related by (\fleq{aeq:povm}):
\begin{subequations}
\begin{align} \label{eq:pvm0}
\ket{\mathcal{A}}_t &= \tiny{\sqrt{2/5}} \qty(\ket{\tilde{0}} + \tiny{\sqrt{3/2}} e^{-i t} \ket{\tilde{1}}), \\
\ket{\chi}_{t}	&= \tiny{\sqrt{1/2}}  \qty( \ket{\tilde{0}} +  e^{\mi t}  \ket{\tilde{1}}).
\end{align}
\end{subequations}
To quantify the effects of noise on projective measurements, we analyze measurements performed on both 3-chain and 5-star cluster states. To isolate the impact of ancillary state noise, we consider ideal cluster states generated via the CZ-gate. Due to cavity losses, the ancilla becomes a mixed state and the resulting post-measurement state is given by \fleq{eq:pvm2}, while the success probability follows \fleq{eq:pvm1}. The results in Table.~\ref{tab:sol} demonstrate identical performance for both geometries across different cavity losses $\beta$. Similar behavior is observed when varying both the projection parameter $t$ and the measurement location.
Measurements in the $XY$-plane suffice for MBQC \cite{mantri2017}, although a $Z$-measurement can be obtained by single atom-cavity iteration. By preparing the atom in $\ket{A_+}$ and the operation $\hat{O}(\pi/2,H,m)$ gives the Kraus operators for measurements $m=\qty{g,s}$ as 
\begin{align}
	K_g	= (Z+I)/2 ~\text{and}~ K_s= (Z-I)/2.
\end{align}

\paragraph{Conclusion and future:}
We analyzed binomial codes and proposed schemes for key components of MBQC: conditional generation of superposition states, deterministic CZ-gate implementation, deterministic cluster state generation, and conditional Pauli measurements in the binomial code space. Atom-cavity and atom-waveguide systems, with coupling strengths ranging from MHz to GHz regimes \cite{2015Reiserer,2018_Chang}, provide an ideal testing ground for these components. Specifically, atom-cavity platforms used for cat-state generation \cite{Hacker_2019} and cluster state generation \cite{Thomas_2022}, along with solid-state setups demonstrating nonlinear phase shifts \cite{Staunstrup2024}, are particularly well-suited for implementing the proposed toolkit.

Recent advancements in atomic trapping and control in 3D architectures \cite{2022_Bluvstein}, enable new possibilities for hybrid quantum computation \cite{Thomas_2024}. The CZ-gate among photons  is implemented via atom-light CZ operation ( Fig.\ref{cz}). This approach naturally extends to the creation of hybrid cluster states combining atomic and photonic qubits \cite{Thomas_2022}. For instance in Fig.\ref{cube}, the face-center qubits can be replaced with atomic qubits, significantly simplifying the circuit to require only atom-cavity reflections to generate hybrid cluster states. Computation then proceeds through the standard MBQC protocol, with atomic and photonic measurements, maintaining fault tolerance and universality while leveraging the complementary advantages of both physical platforms.

Future work could explore deterministic code generation \cite{2025_Kikura}, autonomous error correction under photon loss \cite{Ma_2020}, and POVM implementations based on homodyne detection to simplify resources. However, these are not required for initial experimental tests of the key milestones—state generation, CZ-gates, and cluster state creation—which are readily accessible with current cavity QED technology \cite{Hacker_2019,Thomas_2022, Staunstrup2024}.

\begin{acknowledgments}
	We acknowledge S.M. Girvin for the feedback and Chandan Kumar for discussions. G.P.T acknowledges project CZ.02.01.01/00/22\_008/0004649 (QUEENTEC) and 8C24003 (CLUSSTAR) of MEYS in the Czech Republic, and the EU under Grant Agreement No. 731473 and 101017733 (QuantERA) and also EU project No. 101080173 (CLUSTEC). R.F. acknowledges the Project No.21-13265X of the Czech Science Foundation.
\end{acknowledgments}

\appendix
\subsection{Preparation of superposition states}\label{app:state}
\begin{figure*}[!tbh]
	\subfigure[$\theta = \pi/4 $ \label{p4}]{\includegraphics[scale = 0.36]{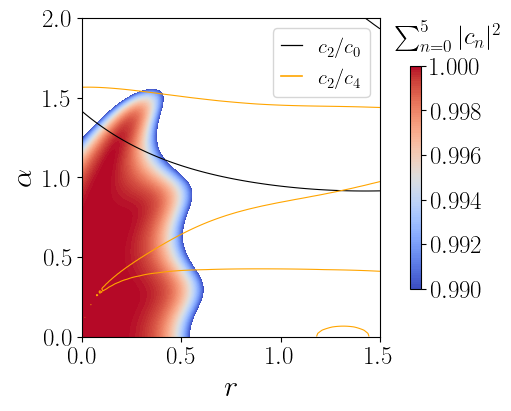}}
	\subfigure[$\theta = \pi/3.3$ \label{p33}]{\includegraphics[scale = 0.36]{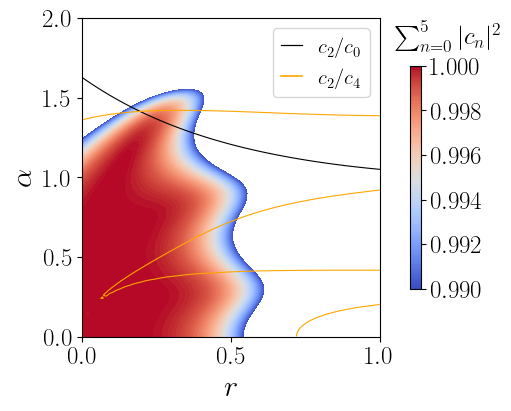}}\quad
	\subfigure[$\theta = \Phi = \pi/4$ \label{af:sol}]{\includegraphics[scale=0.42]{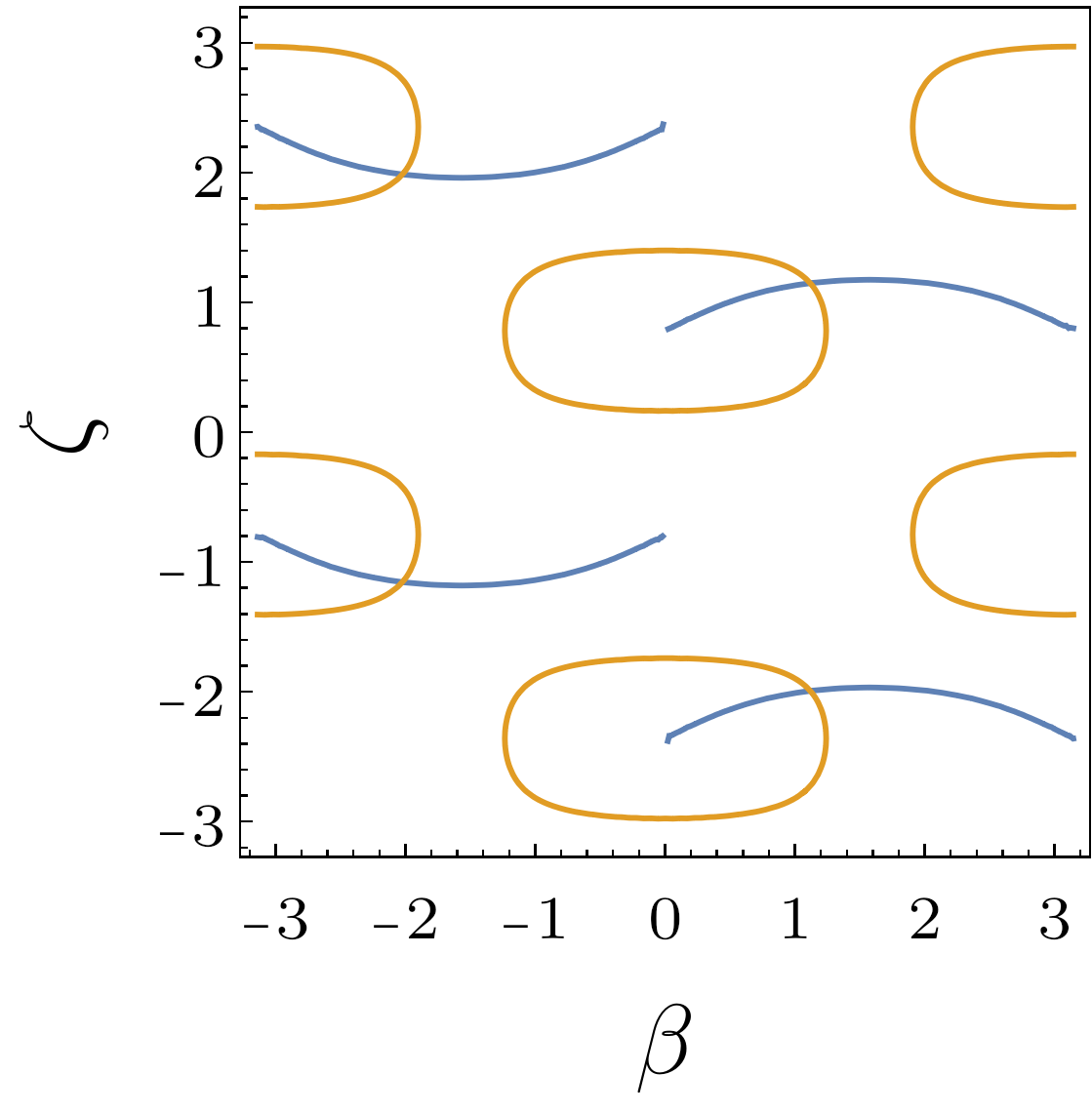}} 
	\subfigure[{$\theta =\pi/8,\Phi = 0$} \label{af:sol1}] {\includegraphics[scale=0.42]{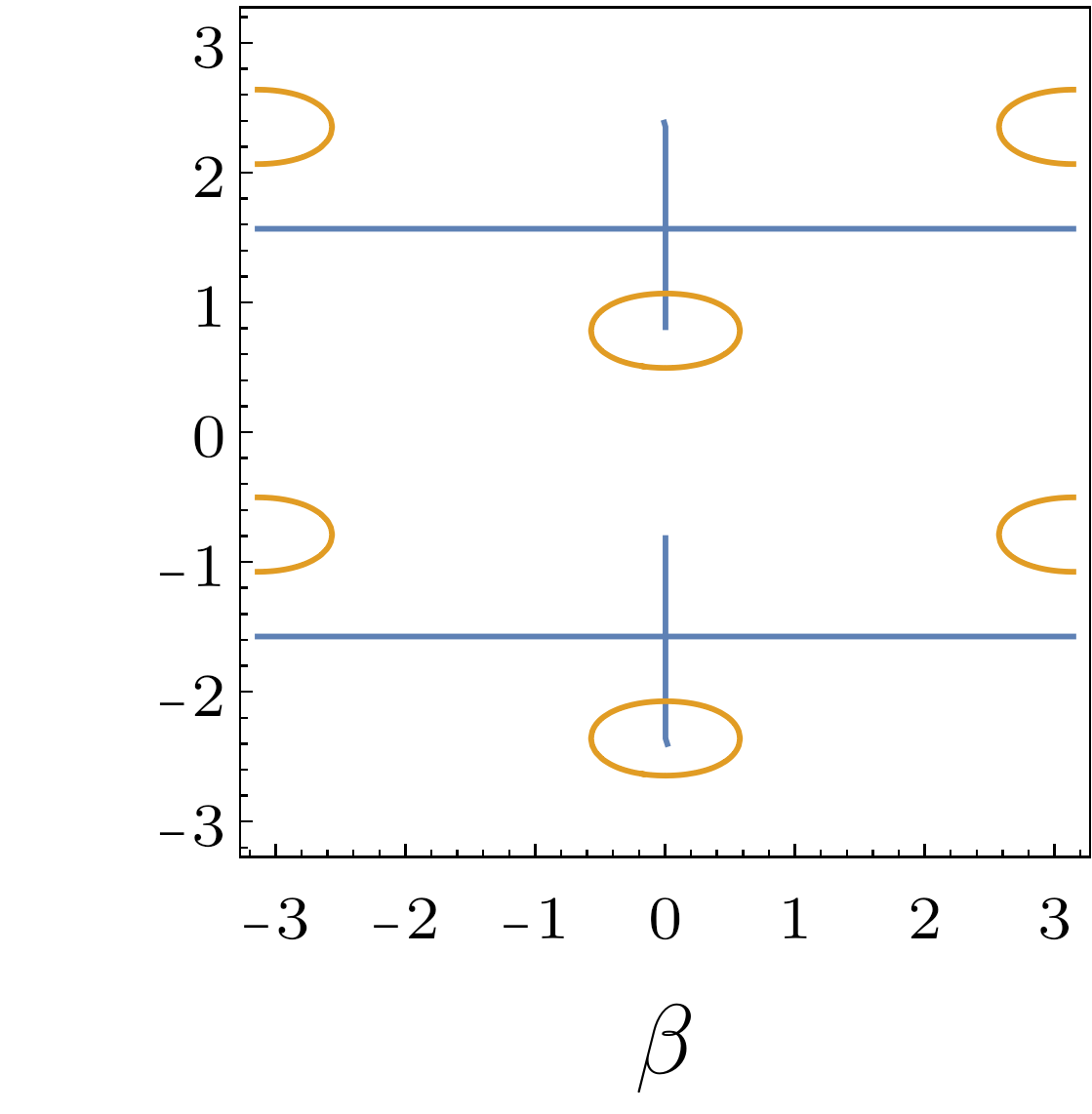}} 
\begin{minipage}{1\textwidth}
\subfigure[Numerical sols\label{tab:sol}]{\raisebox{1.5cm} {
\renewcommand{\arraystretch}{1.5} 
\setlength{\arraycolsep}{1.8pt} 
\resizebox{0.2\columnwidth}{!}{%
\begin{tabular}{w{c}{0.8cm}|c|c|c|c}
state &  $\theta$  &  $\Phi$ & $\beta$ & $\zeta$ \\
\colrule \colrule
			$\ket{+}$	&  $\pi/4$ &  0 &  0 &  -1.714 \\	
			$\ket{T}_1$	& $ \pi/4$ &  $\pi/4$ & 1.174 & -1.979   \\
			$\ket{T}_2$	& T &  $\pi/4$ & 0.575 & -2.103  \\
			$\ket{H}$	&  $\pi/8$ &  0 & 0 & -2.046 \\
			$\ket{\mathcal{A}}$ &   - &  -$\pi/3$  & -1.415 & -2.091				
		\end{tabular}
}}}	\hfill 
\subfigure[Feed forward\label{fg:feed}] {\raisebox{0cm} {\includegraphics[scale=0.28]{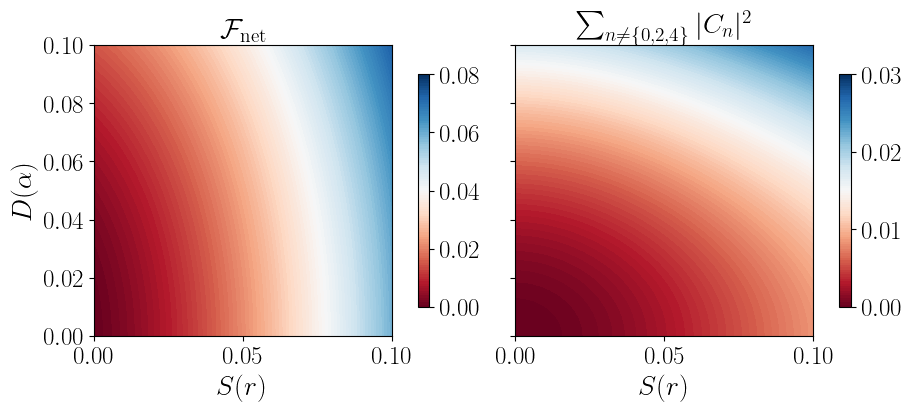}} }\hfill 
\subfigure[Density mat\label{cz_st}]   {\raisebox{0cm} {\includegraphics[scale=0.27]{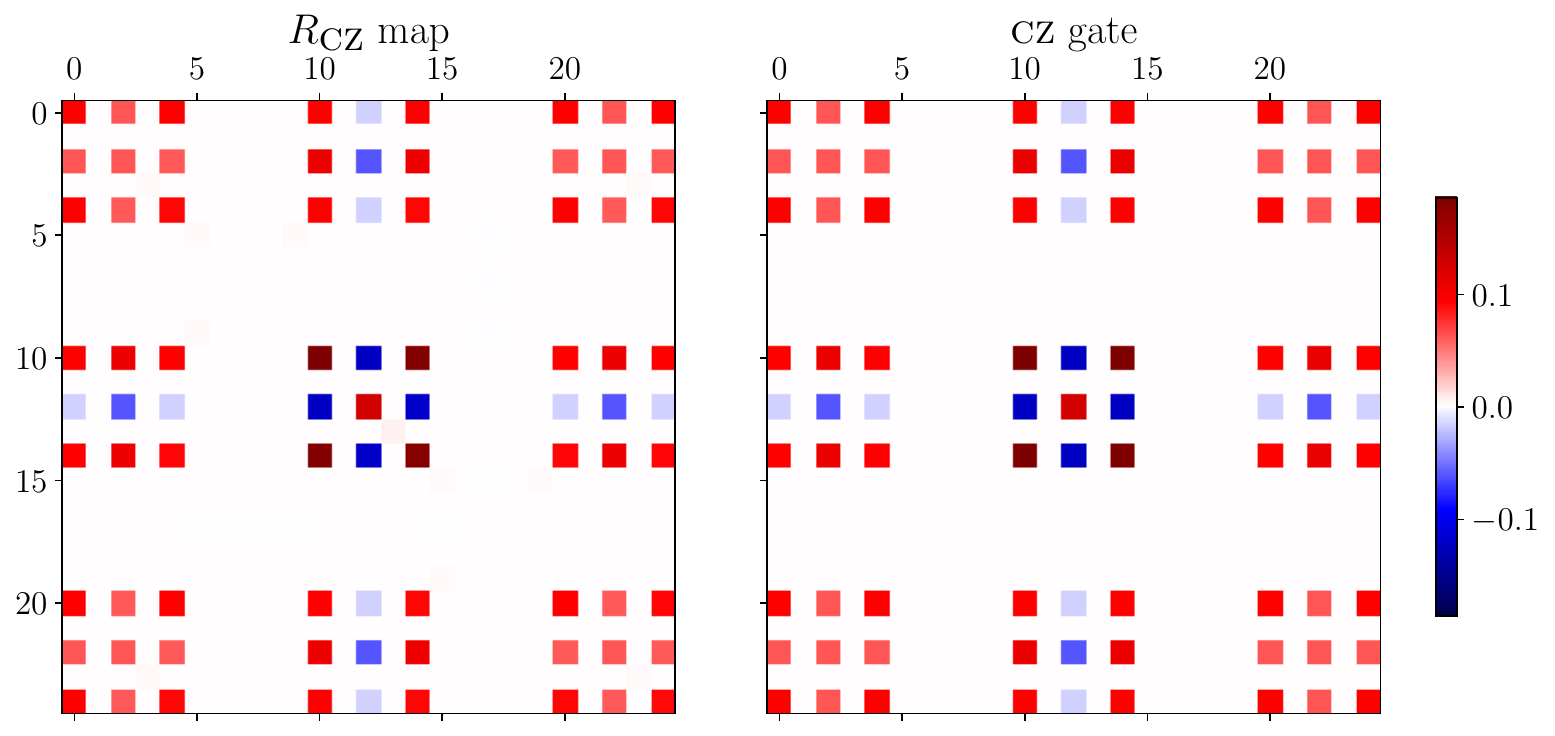}} } 	
\end{minipage}
\caption{\subref{p4} and \subref{p33}: Optimization of amplitude ratios versus higher-order Fock state contributions for squeezed-displaced input states. 
It is clear that the state $\ket{+} = \ket{B}_{(\frac{\pi}{4},0)}$ is not viable through input state optimization and only the state $\ket{\mathrm{B}}_{(\frac{\pi}{3.3},0)}$ is achievable. 
\subref{af:sol} and \subref{af:sol1}: contour plots of \fleq{eq:arn} for $\ket{T}_1$ and $\ket{H}$ states. Table.\ref{tab:sol} shows the $\qty{\beta, \zeta}$ corresponding to $\qty{\theta, \Phi}$ obtained from similar plots. 
The $\theta$ for $\ket{T}_2$ is defined as $\cos(2 \text{T}) = 1/\sqrt{3}$.
\subref{fg:feed} $\mathcal{F}_{\text{net}} \geq 0$ along with $\sum_{n \notin \{0,2,4\}} |C_n|^2$ demonstrate that the fidelity, after the atom-cavity iteration exceeds the fidelity with Gaussian operations.
\subref{cz_st} Comparison of density matrices from $R_\text{CZ}$ with ideal CZ gate for $\beta=0.999$, here we use  $\psi_1= \cos(\pi/4) \ket{\tilde{1}} + e^{\mi \pi/4} \sin(\pi/4) \ket{\tilde{0}}$ and $\psi_2 = \cos(\pi/3) \ket{\tilde{1}} + e^{\mi \pi/5} \sin(\pi/3) \ket{\tilde{0}}$  and the fidelity with ideal gate operation is 0.981.}\label{fg:ap1}
\end{figure*}
To prepare a general superposition state of binomial codes, we begin by defining a unitary matrix in the basis $\ket{g}=\tiny\mqty(1\\0)$ and $\ket{s}=\tiny\mqty(0\\1)$:
\begin{align}
R[\mu,\beta,\zeta] & = \mqty[
e^{\mi \mu}	\cos \zeta & e^{\mi \beta} \sin \zeta\\
-e^{-\mi \beta} \sin \zeta &  e^{-\mi \mu}\cos \zeta	
],\label{app:unit}
\end{align}
note that the Hadamard is given as $H=R[\pi/2,\pi/2,\pi/4]$. Also we denote $R[0,\beta,\zeta] \equiv R[\beta,\zeta]$ and $R[0,0,\zeta] \equiv R[\zeta]$.

An atom-cavity iteration without measurement $\hat{O}(\pi/2,H,I)$ transforms 
the state $\ket{l} \otimes \ket{A_+}$ as
\begin{align}
\rho^{(1)} = \frac{1}{2}\qty[ (\bar{\ket{l}} + \ket{l}) \ket{g} + (\bar{\ket{l}} - \ket{l}) \ket{s} ],
\end{align}
where $\ket{\bar{l}} = e^{i \frac{\pi}{2} \hat{n}} \ket{l}$. Tracing out the atomic degree of freedom yields the mixed state $\frac{1}{2}\qty[\dyad{\bar{l}}+\dyad{l}]$. For instance, setting the input ($\ket{l}$) to a coherent state gives the state after first iteration as $\rho^{(1)} = \frac{1}{2}\qty[\dyad{-l}+\dyad{l}]$, which differs from the cat state $\frac{1}{\sqrt{2}}\qty[ \ket{-l} \pm \ket{l}]$ obtained with atomic measurement. 
Instead of a coherent state, using optimized squeezing and displacement parameters $\qty{\alpha, r}$ produces $\ket{\mathrm{B}}_{\pi/3.3,0}$, Fig.~\ref{p33} . A second atom-cavity reflection gives 
\begin{widetext}
\begin{equation}
U({\pi}/{2}) \ket{\mathrm{B}}_{\pi/3.3,0} = \qty[\cos(\frac{\pi}{3.4}) \ket{\tilde{0}} - \sin(\frac{\pi}{3.4}) \ket{\tilde{1}}] \ket{g} +
	\qty[\cos(\frac{\pi}{3.4}) \ket{\tilde{0}} + \sin(\frac{\pi}{3.4}) \ket{\tilde{1}}] \ket{s}
\end{equation}
then the atomic-rotation  $R(\alpha,\beta,\theta)$ further transform the state to
\begin{align*}
\begin{aligned}
[e^{\mi \mu} \cos\zeta \cos ({\pi}/{3.4}) + e^{\mi \beta}\sin\zeta\cos ({\pi}/{3.4})] \ket{\tilde{0}}  +
		 [- e^{\mi \mu} \cos\zeta \sin ({\pi}/{3.4})+ e^{\mi \beta}\sin\zeta\sin ({\pi}/{3.4})] \ket{\tilde{1}} \otimes \ket{g}+\\
[-e^{-\mi \beta}\sin\zeta \cos ({\pi}/{3.4}) + e^{-\mi \mu} \cos\zeta \cos({\pi}/{3.4})] \ket{\tilde{0}} + 
[e^{-\mi \beta}\sin\zeta \sin ({\pi}/{3.4})+ e^{-\mi \mu}\cos\zeta\sin ({\pi}/{3.4})] \ket{\tilde{1}} \otimes \ket{s}.
\end{aligned}
\end{align*}
Now with $\mu=0$, the atomic measurements projects the light field to the state
\begin{align}\label{ape:frn}
\ket{\tilde{0}}  +\dfrac{-\cos\zeta + e^{\mi\beta}\sin\zeta}{\cos\zeta + e^{\mi \beta}\sin\zeta} \tan(\frac{\pi}{3.4}) \ket{\tilde{1}} \text{for} \ket{g}  ~\text{and}~
\ket{\tilde{0}}  + \dfrac{\cos\zeta + e^{-\mi \beta} \sin \zeta }{\cos\zeta - e^{-\mi \beta}\sin\zeta} \tan(\frac{\pi}{3.4})  \ket{\tilde{1}}  ~ ~\text{for} ~ \ket{s} ,
\end{align}
and finally we optimize $\qty{\zeta,\beta}$ to match the ratio to a desired superposition
$
\dfrac{-\cos\zeta + e^{\mi \beta} \sin \zeta }{\cos\zeta + e^{\mi \beta}\sin\zeta} \tan(\frac{\pi}{3.3}) = \tan \theta ~e^{\mi\Phi}.\\
$
\end{widetext}
In Figs.~\ref{af:sol} and \ref{af:sol1} we show numerical solutions for $\ket{+}$ and $\ket{H}$ superposition states. The Table.~\ref{tab:sol} shows numerical solutions for various $\qty{\theta,\Phi}$.
Note that any general superposition of  $\ket{\tilde{0}} $ and $\ket{\tilde{1}}$ can be filtered as long as \fleq{ape:frn} is solvable.
Although $\ket{\mathrm{B}}_{\pi/3.3,0} \approx \ket{+}$, a second iteration is necessary. This is shown by comparing the fidelities after the first atom-cavity iteration, with and without subsequent Gaussian operations :
\begin{align}
\mathcal{F}_{\text{net}} \equiv \braket{+}{\mathrm{B}}_{\pi/3.3,0} -\mel{+}{\hat{D}(\alpha) \hat{S}(r)}{\mathrm{B}}_{\pi/3.3,0}.
\end{align}
From  Fig.~\ref{fg:feed} we see $\mathcal{F}_{\text{net}} \geq 0$, hence Gaussian operations are not a substitute for atom-cavity iteration. The fidelity reduction arises because squeezing and displacement introduce undesirable higher and odd photon-number components, quantified by $\sum_{n \notin \{0,2,4\}} |C_n|^2$, where $C_n$ are the amplitudes of $\hat{D}(\alpha)\hat{S}(r)\ket{\mathrm{B}}_{\pi/3.3,0} $.

\section{Process map ($R_\text{CZ}$)}
To realize the CZ-gate among qubits encoded in binomial codes, we first rewrite 
$e^{\mi \frac{\pi}{2} \hat{n} }$ as
$
\qty[\dyad{\tilde{0}}+\dyad{\tilde{1}}]~
e^{\mi \frac{\pi}{2} \hat{n} }~ \qty[\dyad{\tilde{0}}+\dyad{\tilde{1}}] = Z
$
which modifies $U(\frac{\pi}{2})$ to 
$
U = Z \dyad{g} + I \dyad{s},
$
\begin{itemize}
\item Reflecting the two light modes will give total operation:
$U \otimes U = Z Z   \dyad{g} + I  I \dyad{s}$.
\item Initializing the atom in $\ket{A_+}$ and ${H}$-rotation on atom gives: 
$(Z  Z + I I) \ket{g} + (Z Z - I I) \ket{s}$.
\item Applying $U$ only to first light mode gives:\\
$(I  Z  +   Z I ) \ket{g} + ( Z  Z - II) \ket{s}$.
\item $\bar{H}$-rotation on atom gives:\\
$=(Z  I + I Z-Z  Z + I I)\ket{g} + (Z  I + I Z + Z  Z - I  I)\ket{s} \\
=\text{diag}(1,1,1,-1)	\ket{g} + \text{diag}(1,-1,-1,-1)\ket{s}.
$
\end{itemize}
The $R_\text{CZ}$ map is obtained directly when measuring the $\ket{g}$ state. When the measurement is  $\ket{s}$, we apply the phase operation $e^{i\frac{\pi}{2}\hat{n}}$ to both modes i.e. diag(1,-1,-1,1), rendering the  operation deterministic.

To assess the quality of the $R_\text{CZ}$ map as CZ-gate, we first note that the two-qubit Pauli matrices 
\begin{align}
\sigma_k \in \qty{II, IX, IY, IZ, XI, XX, \dots, ZZ},
\end{align}
form an orthonormal basis for  $4 \times 4$ matrices. A matrix $\rho$ can be vectorized as $V_k = \tr[\sigma_k \rho]$ 
and the process map $R_\text{CZ}$ is vectorized to 
\begin{align}
 \ket{b}^j  = \tr[\sigma_k \rho_\text{out}^j] = \tr[\sigma_k R_\text{CZ} (\rho_\text{in}^j) ],
\end{align}
note that the Pauli basis $\sigma_k$ matrices  are not valid density matrices, we therefore construct density matrices using linearly independent projectors \cite{2019_Hu,Chou_2018}
\begin{align}\label{eq:bss}
\begin{aligned}
\lambda_j \in   \{ \ket{\tilde{0}}, \ket{B}_{(\frac{\pi}{4},0)}, \ket{B}_{(\frac{\pi}{4},\frac{\pi}{2})}, \ket{\tilde{1}} \}^{\otimes 2},\\
\ket{a}^j  = \tr[\sigma_k \rho_\text{inp}^j]  ~\text{with}~ \rho_\text{inp}^j = \dyad{ \lambda_j}.
\end{aligned}
\end{align}
Then the map $R_\text{CZ}$ is written as:
\begin{align}\label{eq:dcz}
\qty[\ket{b}^1, \ket{b}^2\dots\ket{b}^{16}] &= R_\text{CZ} \qty[\ket{a}^1, \ket{a}^2\dots\ket{a}^{16}],  \nonumber \\
R_\text{CZ} &= [b] [a]^{-1}.
\end{align}
Fig.~\ref{cz1} compares the ideal CZ-gate and with $R_\text{CZ}$-gate using atomic cavity iterations,
similarly  Fig.~\ref{cz_st} shows the density matrices after ideal-gate  and  cavity-gate operation.

\section{Projective measurements}
Measurements in the $XY$-plane i.e., eigenstates of $\cos t ~\hat{X} + \sin t ~\hat{Y}$, can be performed using the circuit in Fig.~\ref{mc}. The POVM and the success probability corresponding to detecting $\{n_a, n_b\}$ photons is given by \cite{2009_Garciana}
\begin{align}\label{eq:pvm}
\ket{\chi}  &= \dfrac{1}{\sqrt{n_a! n_b!}}\mel{\mathcal{A}_a}{ ~U^\dagger \hat{a}^{\dagger n_a} \hat{b}^{\dagger n_b}   U  ~} {0_a,0_b},\\
 \mathcal{P} & =  \text{Tr} [ \rho_{ab} { U^\dagger} \dyad{n_a,n_b} U] \label{eq:pvm1}
\end{align}
here $U = e^{\mi \frac{\pi}{4} (\hat{a}^\dagger \hat{b} + \hat{a} \hat{b}^\dagger)}$ is the BS operation and $\rho_{ab}$ is the total density matrix of ancilla and input state. The $\ket{\chi}$  for detecting two photons in each mode  ($n_a=n_b=2$) simplifies \fleq{eq:pvm}:
\begin{align}
\ket{\chi} = {1}/{8}\mel{\mathcal{A}_a}{  (\hat{a}^{\dagger } + \mi \hat{b}^{\dagger})^2  (\hat{a}^{\dagger } - \mi \hat{b}^{\dagger})^2} {0_a,0_b},
\end{align}
Typically, the auxiliary state is a coherent state but replacing it with an optimized non-classical state $\ket{\mathcal{A}}_t = \cos\theta \frac{ \ket{0}+ \ket{4} }{\sqrt{2}}  + \sin\theta e^{\mi t} \ket{2} $ gives
\begin{align}\label{aeq:povm}
\ket{\chi}_t = \dfrac{\cos\theta  \sqrt{4!}  }{8} \qty(   \qty[\frac{ \ket{0}+ \ket{4} }{\sqrt{2}}]  + \sqrt{\frac{2}{3}} e^{-\mi t} \tan\theta \ket{2} ),
\end{align}
then the required projection $\ket{\chi}_t = \frac{1}{\sqrt{2}}\qty(\frac{\ket{0}+ \ket{4}}{\sqrt{2}} + e^{-\mi t} \ket{2} )$ is realized by adjusting $\sqrt{\frac{2}{3}} \tan\theta =  1$ and the success probability is given by $\text{Tr}[\dyad{\chi} \rho]$. further  \fleq{eq:pvm} can be generalized to a mixed state of ancilla: 
\begin{align}\label{eq:pvm2}
\begin{aligned}
\tilde{\rho}_b  &=   \rho_b \text{Tr}_a \qty(\rho_a U^\dagger \dyad{2,2} U),\\ 
& =  \sum_a \lambda_a  \dyad{\chi_a} \rho_b \dyad{\chi_a},
\end{aligned}
\end{align}
where $\lambda_a$ and $\chi_a$ are eigenvalues and eigenvectors of $\text{Tr}_a \qty(\rho_a U^\dagger \dyad{2,2} U)$. Note that $\sum_a \lambda_a  \dyad{\chi_a} < 1$ as it corresponds only to detecting 2-photons.

\end{document}